# Frequency content and filtering of head sensor kinematics: A method to enable field-based inter-study comparisons

Gregory Tierney,[1,2] Steve Rowson,[3] Ryan Gellner,[3] Sadaf Iqbal,[1] Pardis Biglarbeigi,[4] James Tooby,[2] James Woodward,[1] Amir Farokh Payam[5]

[1] Sport and Exercise Sciences Research Institute, Ulster University, Belfast, United Kingdom

[2] Carnegie Applied Rugby Research (CARR) Centre, Carnegie School of Sport, Leeds Beckett University, Leeds, United Kingdom

[3] Biomedical Engineering and Mechanics, Virginia Tech, Blacksburg, Virginia, USA

[4] School of Science and Engineering, University of Dundee, Dundee, United Kingdom

[5] Nanotechnology and Integrated Bioengineering Centre (NIBEC), Ulster University, Belfast, United Kingdom

**Corresponding author:**

Name: Dr Gregory Tierney     Email: g.tierney@ulster.ac.uk

Address: Sport and Exercise Sciences Research Institute, Ulster University, Belfast, UK

## ABSTRACT

Wearable head sensor systems use different kinematic signal processing approaches which limits field-based inter-study comparisons, especially when artefacts are present in the signal. The aim of this study is to assess the frequency content and characteristics of head kinematic signals from head impact reconstruction laboratory and field-based environments to develop an artefact attenuation filtering method (artefact attenuation method). Laboratory impacts ($n=72$) on a test-dummy headform ranging from 25-150 g were conducted and 126 elite-level rugby union players were equipped with instrumented mouthguards (iMG) for up to four matches. Power spectral density (PSD) characteristics of the laboratory impacts and on-field HAE ($n=5694$) such as the 95th percentile cumulative sum PSD frequency were utilised to develop the artefact attenuation method. The artefact attenuation method was compared to two other common filtering approaches (Fourth order (2x2 pole), zero-lag Butterworth filter with 200 Hz (-6 dB) cut-off frequency (Butterworth-200Hz) and CFC180 filter) through signal-to-noise ratio (SNR) and mixed linear effects models for laboratory and on-field events, respectively. The artefact attenuation method produced an overall higher SNR than the Butterworth-200Hz and CFC180 filter and on-field peak linear acceleration (PLA) and peak angular acceleration (PAA) values within the magnitude range tested in the laboratory. Median PLA and PAA were higher for the CFC180 filter than the Butterworth-200Hz ($p<0.01$) and artefact attenuation method ($p<0.01$), reporting values as high as 294 g and 31.2 krad/s$^2$. The artefact attenuation method can be applied to all commercially available iMG kinematic signals with adequate sample rates to enable field-based inter-study comparisons.

## INTRODUCTION

The collision nature of many contact sports means concussion and repetitive head acceleration events (HAE) are an issue.[1-3] HAE can occur directly and indirectly from head and body contact on the field, respectively.[1,4] Head kinematics are associated with brain injury risk with rotational head motion considered the primary contributor to brain injury.[1] A recent review article indicates a growing evidence base of various biomechanical brain injury mechanisms, including those involving repetitive HAE.[1] Repetitive HAE will continue to be a concern in contact sports until proactive mitigation strategies are developed and this requires an understanding of the biomechanical mechanisms.

Historically, many studies on HAE in sport have been limited by wearable head sensor validity that suffer from soft tissue artefacts (e.g., instrumented headgear and patches).[5] Wearable head sensors with superior coupling to the skull provide a unique opportunity for measuring HAE kinematics in sport (e.g., instrumented mouthguards (iMG) and mouthpieces).[5] Accordingly, player head acceleration exposure and the biomechanical mechanisms of HAE and concussion in sport can be further understood. A recent study on iMG validity found that the majority of iMG devices performed highly in a head impact reconstruction laboratory.[6] However, the study illustrated that on-field head kinematics reported by iMG differed considerably with certain systems producing much higher head kinematics (roughly 500 g and > 50 krad/s$^2$) in comparison to others. At present, each iMG system uses different kinematic signal processing approaches (e.g., filter cut-off frequencies, machine learning-based noise/artefact detection models) which limits cross-study comparison.[1,6,7] Wu et al.[8] illustrated through cadaver head impact reconstructions that low-pass filter cut-off frequencies (-3 dB) of 590 and 290 Hz, were required to keep amplitude attenuation within 10% for linear acceleration and angular velocity, respectively. However, the highest -3 dB cut-off frequency used by an iMG system in Jones et al.[6] was 300 Hz, through the use of a Channel Frequency Class 180 (CFC180) filter.

A head impact reconstruction laboratory provides an idealised environment for testing iMG.[1,6,7,9,10] The impact magnitudes are controlled to a high degree through the use of an impactor that strikes a test-dummy head-neck model, and the iMG is clamped onto a plastic upper dentition within the headform. [1,6,7,9,10] During these impacts, the processing of noise is straightforward and/or standardised.[9] Sources of noise in these reconstructions can occur from the measurement system (i.e. headform sensor array), numerical differentiation (if applicable) and biomechanical transformations of this data (e.g., transforming the linear acceleration signal from the sensors to the head centre of gravity (CG)).[11] Performing well in a head impact reconstruction laboratory does not constitute a valid device for use in the field.[9] It provides a basic check that the hardware can measure impacts we expect to see on the field and the firmware/software can process and transform the data appropriately.[9]

In real world environments such as the rugby field, there are cases where artefacts will exist in the iMG signal.[4,12] Example sources of artefacts in the context of iMG include movement of the mouthguard on the teeth due to poor fit/coupling, shouting, biting, mandible interference and direct impacts to the mouthguard.[4] Video-verification based on-field validation to quantify true-positive, false-positive and false-negative metrics can give an indication of the iMG system HAE detection algorithms.[7] However, true-positive events can have artefacts in the signal if left unattenuated. These artefacts have the potential to produce unwanted frequency content in the signal.[12] Accordingly, artefacts can result in erroneous head kinematics being reported by the iMG system which is further compounded when numerical calculations and biomechanical transformations (e.g., transforming linear acceleration signal to the head CG) are implemented.[11,12]

A head impacting a rigid surface as a perfectly elastic rigid body detached from the neck can be modelled as a spring-mass system (Supplementary A). The linear acceleration of the head during the impact is described by Equation 1 and derived in Supplementary A.[13]

$$a(t) = -v_i\sqrt{\frac{k}{m}}sin\,(\sqrt{\frac{k}{m}}t) \qquad [1]$$

Where $a$ is the linear acceleration of the head, $v_i$ is head velocity at initial time of impact, $k$ is the head stiffness, $m$ is the head mass and $t$ is time. Equation 1 illustrates that the acceleration of the head during an impact has 1) a characteristic pulse modelled as a half-sine wave and 2) amplitude and pulse duration influenced by the impact conditions. That is, the magnitude of the acceleration is dependent on the initial head velocity, mass and stiffness and the acceleration pulse duration (or frequency) is dependent on the head mass and stiffness.

The dynamic nature of contact sports means a range of impact conditions are likely to occur on the field during HAE.[1] Sports collisions can include different impact velocities, effective masses and complex contact characteristic (e.g., non-linear stiffness and damping characteristics, influence of neck and body).[14,15] Head impact reconstruction laboratories provide the ability to test to extreme HAE magnitudes and durations seen on the sports field in a controlled environment with artefacts mitigated.[6,9] The aim of this study is to assess the frequency content and characteristics of head kinematic signals from head impact reconstruction laboratory and field-based environments. An artefact attenuation filter will be developed based on the findings and applied to an elite-level rugby union instrumented mouthguard dataset. The aim of the artefact attenuation filter is to enable inter-study comparisons for field-based iMG and instrumented mouthpiece studies.

## MATERIALS AND METHODS

### *Head impact reconstruction laboratory test set-up*

The dummy headform configuration comprises of a medium-sized National Operating Committee on Standards for Athletic Equipment (NOCSAE) headform affixed to a Hybrid III 50th percentile male neck, fitted on a linear slide table with 5 degrees of freedom.[6,9] Reference kinematics were measured at the dummy headform CG with an instrumentation package comprising of three linear accelerometers (Endevco 7264b-2000; Meggitt Orange County, Irvine, CA) and a tri-axial angular rate sensor (DTS ARS3 Pro 18k; Diversified Technical Systems, Seal Beach, CA), both recording at 20 kHz.[6,9] No filter was applied to the signals apart from the default hardware filters (-3 dB cut-off frequency of 4000 Hz). Head impact laboratory reconstruction testing was conducted utilising a pendulum impactor to simulate bareheaded impacts to the dummy headform. Tests captured the range of HAE magnitudes and durations seen in rugby through the use of a rigid (nylon, 25 mm thickness) and padded (vinyl-nitrile foam, 40 mm thickness) impactor (127 mm diameter) to the bareheaded dummy headform. Impacts occurred at the front and front boss locations of the headform at target linear head accelerations of 25, 50, 75, 100, 125 and 150 g. Three tests were conducted at each configuration, totalling 72 impacts (6 impact magnitudes x 2 impactors x 2 locations x 3 tests per configuration).

*On-field instrumented mouthguard data collection*

A total of 126 participants (96 male and 30 female) were recruited from seven (four male and 3 female) elite-level rugby union club/international teams in the northern hemisphere. Each participant provided written consent and ethical approval for the study was given by the University's Research Ethics Committee (UREC), University of Ulster (#REC-21-0061). Participants underwent 3D dental scans and were provided with a custom-fit iMG (Prevent Biometrics, Minneapolis MN). The iMG were equipped with an accelerometer and gyroscope both sampling at 3200 Hz and a measurement range of ± 200 g and ± 35 rad/s, respectively.[6,16] Laboratory and on-field validity of the Prevent Biometrics iMG has been demonstrated in previous studies.[6,9,10] Most recently, a validity and feasibility study of multiple iMG systems illustrated that the Prevent Biometrics iMG had a concordance correlation coefficient value of 0.984 in laboratory-based impact testing using a crash test dummy headform.[6,7] The trigger mechanism for the iMG in this study was when 8 g was exceeded on a single axis of the iMG accelerometer, capturing 10 ms of pre-trigger data and 40 ms of post-trigger data. All linear kinematics were transformed from the iMG to the head CG and a 5 g recording threshold was applied. Peak linear acceleration (PLA) and peak angular acceleration (PAA) of the head were extracted from each HAE.

The level of noise/artefact in the kinematic signal was classified by an in-house Prevent Biometrics classification algorithm that determined whether each HAE contained minimal noise (class 0), moderate noise (class 1) or severe noise (class 2). A fourth order (2x2 pole) zero phase, low-pass Butterworth filter was applied to each signal with a cut-off frequency (-6 dB) of 200, 100 and 50 Hz for class 0, 1 and 2 HAE. A -6 dB cut-off frequency is a consequence of not adjusting the -3 dB cut-off frequency when filtering twice (once forward and once backward), in order to prevent phase shifts.[11] The filter is applied as follows: 1) Filter applied to raw linear acceleration and angular velocity 2) Filter applied to differentiated angular velocity (i.e. angular acceleration) 3) Filter applied to linear acceleration when transformed to head CG.[17] Data were collected from 209 player-matches (160 male and 49 female). Multiple angle, broadcast quality video footage was available for all matches. HAE were synchronised to the video to a 40 ms resolution enabling video verification. Both the on-field data collection head impact reconstruction laboratory impacts complied with the consensus head acceleration measurement practices (CHAMP).[4,18]

*Artefact attenuation filter method development*

The design of the artefact attenuation filtering approach (artefact attenuation method) predominantly utilised the power spectral density (PSD) characteristics of the head impact reconstruction laboratory impacts.[19] The component (X, Y or Z) of raw linear acceleration signal with the highest peak magnitude was utilised for the PSD analysis. PSD was conducted only on the impact pulse.[20] The impact pulse was calculated based on the first zero-crossing timepoint before and after peak linear acceleration was reached. The frequency associated with the 95th percentile PSD (cumulative sum) was extracted from each laboratory impact.[11,19] Each pulse (on-field and laboratory) was padded using the *nextpow2* function in MATLAB (MathWorks, Inc., Natick, MA). The maximum 95$^{th}$ percentile PSD frequency ($F_{max}$) for the laboratory impacts was considered the upper limit for on-field HAE.

The same PSD approach was conducted for on-field HAE. Any on-field HAE with 95$^{th}$ percentile PSD above $F_{max}$ was considered to have an artefact in the signal. Additionally, the first local minima frequency on the PSD spectrum was extracted from on-field HAE as signal above this was considered to contain unwanted higher-frequency noise due to signal artefacts.[11,21,22] A CFC filter was utilised for the artefact attenuation method based on impact biomechanics standards (i.e., SAE J211).[23] In brief, a CFC filter is a fourth order (2x2 pole), zero phase, low-pass Butterworth filter with frequency class ($F_H$) indicating the entire flat range of the frequency response (i.e. just before roll-off).[23] For a CFC filter, $F_H$ is equal to 0.6 multiplied by the -3 dB cut-off frequency (excluding CFC1000).[23]

If at least one frequency was below $F_{max}$, $F_H$ was selected as the lower frequency associated with the 95th percentile PSD and local minima (if applicable). If neither frequencies were below $F_{max}$, a CFC60 filter (100 Hz -3 dB cut-off frequency) was applied. CFC60 was the lowest CFC filter applied to a kinematic signal unless a signal component exceeded the sensor range, then a CFC30 filter was applied (<0.3% of cases). $F_H$ was applied to all sensor component channels. The analysis was conducted separately on both the linear and angular acceleration pulse to mitigate attenuating signal when artefacts were more prominent in one sensor. An artificial 'raw' angular acceleration pulse was developed by filtering the raw angular velocity with a CFC filter with $F_H = F_{max}$ and then differentiating the signal (five-point stencil method).[24] This approached ensured that a 95th percentile PSD frequency less than $F_{max}$ was not artificially created and that low magnitude events were not corrupted by low amplitude, high frequency noise in the raw angular velocity signal.

*Data and statistical analysis*

A continuous wavelet transform (CWT) was calculated for each laboratory impact and on-field HAE.[25] To assess the performance of the artefact attenuation method in a head impact reconstruction laboratory environment, signal to noise ratio (SNR) was calculated with reference to the raw signal and compared to a fourth order (2x2 pole), zero phase, low-pass Butterworth filter with -6 dB cut-off frequency of 200 Hz (Butterworth-200Hz) and a CFC180 filter applied to the raw headform linear acceleration and angular velocity signals.[26] The Butterworth-200Hz filter was selected to represent the Prevent Biometrics processed output if their in-house classification algorithm was not applied. A CFC180 filtering approach was included as it is the filter type with highest -3dB cut-off frequency (300 Hz) utilised by an iMG system tested by Jones et al.[6]

The peak linear acceleration (PLA) at the head CG and peak angular acceleration (PAA) were extracted from each on-field HAE based on the Prevent Biometrics processed output (i.e., cut-off frequency (-6 dB) of 200, 100 and 50 Hz for class 0, 1 and 2 HAE). Median, Interquartile ranges (IQR) and maximum values for PLA and PAA were compared for: 1) Prevent Biometrics processed output 2) Artefact attenuation method 3) Butterworth-200Hz 4) CFC180 filter through mixed liner effects modelling with an alpha level of $p < 0.05$, which has previously been described in detail.[27,28] The artefact attenuation method, Butterworth-200Hz and CFC filters were applied at the same stages as the Prevent Biometrics processed output. Transforming the linear acceleration signal at the iMG to the head CG was validated by getting the same results as the Prevent Biometrics processed output when using 200 Hz (-6 dB cut-off frequency) for class 0 impacts.

**RESULTS**

The maximum 95$^{th}$ percentile PSD frequency for the laboratory impacts was 312 Hz and was achieved using the rigid impactor at 50-150 g. The pulse duration ranged from 3.2 to 12.3 ms for the laboratory impacts. The acceleration pulse during the laboratory impacts followed a half-sine/haversine shape with the CWT scalogram illustrating a smooth frequency spectrum and PSD exhibiting no local minima deflection points (Fig. 1a-c). Median SNR (and IQR) for the artefact attenuation method and CFC180 filter were similar at 19.6 dB (17.8 to 21.1) and 19.7 dB (17.8 to 21.2), respectively for linear acceleration. Higher median SNR was achieved by the artefact attenuation method (26.1 dB; IQR: 21.9 to 30.0) compared to the CFC180 filter (24.8 dB; IQR: 21.8 to 30.4) for angular velocity. Both the artefact attenuation method and CFC180 filter had greater SNR than the Butterworth-200Hz for linear acceleration (16.4 dB; IQR: 11.0 to 19.1) and angular velocity (-0.3 dB; IQR: -1.0 to -0.1).

A total of 5694 video-verified HAE with raw data signals were collected from the on-field study. The maximum 95$^{th}$ percentile PSD for on-field HAE was 1550 Hz and pulse duration ranged from 0.6 to >50 ms (50 ms was the iMG time window). The raw acceleration pulse during certain on-field HAE did not follow a half-sine or haversine pulse shape with the CWT scalogram illustrating localised areas of high frequency and amplitude signal and PSD exhibiting local minima deflection points (Fig. 1d-f). For the artefact attenuation method, $F_{max}$ was rounded up to 320 Hz (from 312 Hz). $F_H$ was selected from the 95$^{th}$ percentile and first local minima PSD frequency more for linear than angular acceleration pulses (72.9% vs. 62.7% and 21.0% vs. 17.8%, respectively) with 5.8% and 19.2%, respectively, having neither within $F_{max}$, see Fig. 2. The median and IQR for both the linear and angular $F_H$ was 60 Hz (60-100 Hz) with maximum values of 300 Hz.

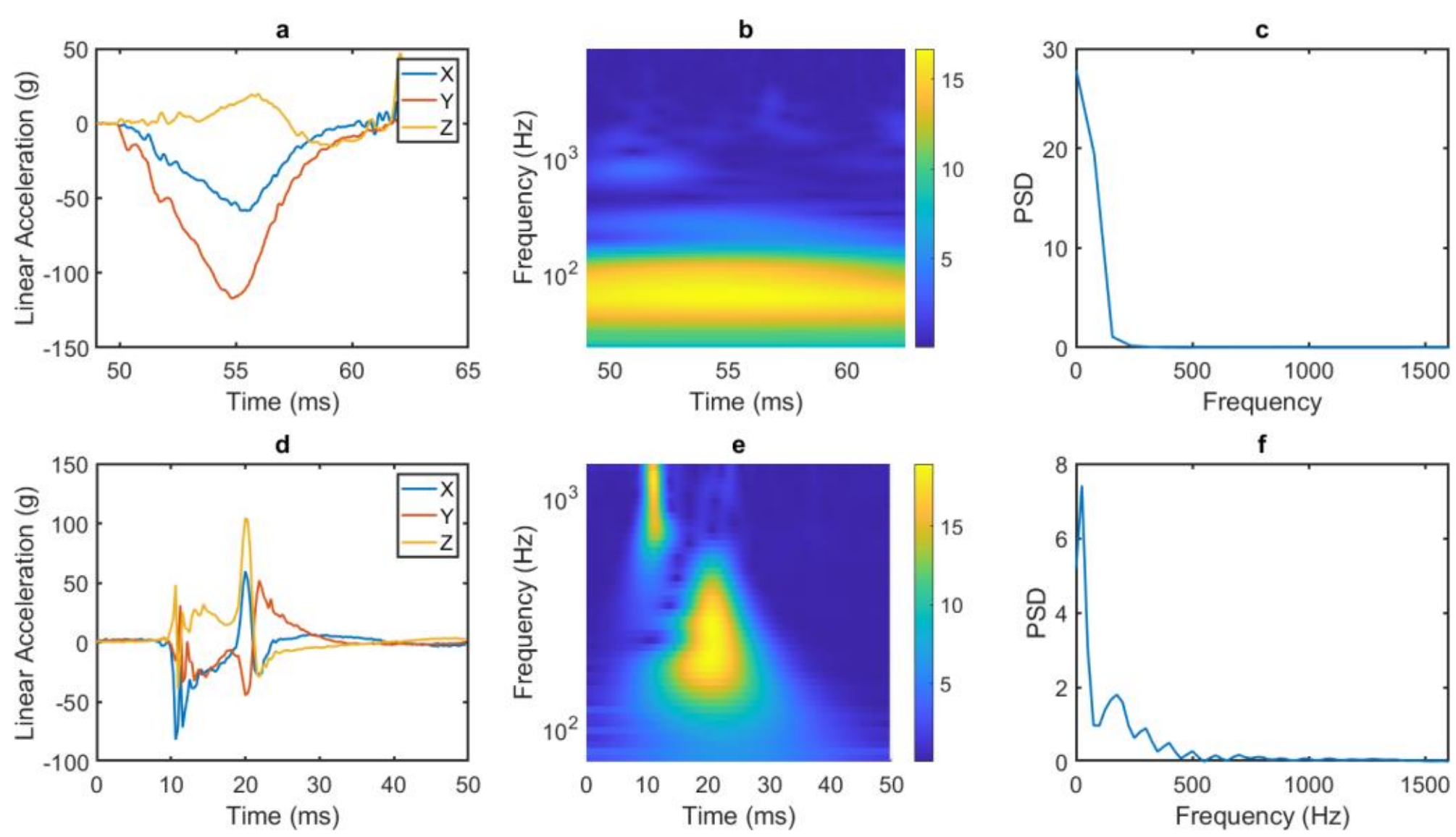

**Figure 1.** Linear acceleration time-series, CWT scalogram (log-scale on y-axis) and PSD for an example laboratory impact (a-c) and on-field HAE with localised area of high frequency and amplitude signal (d-f).

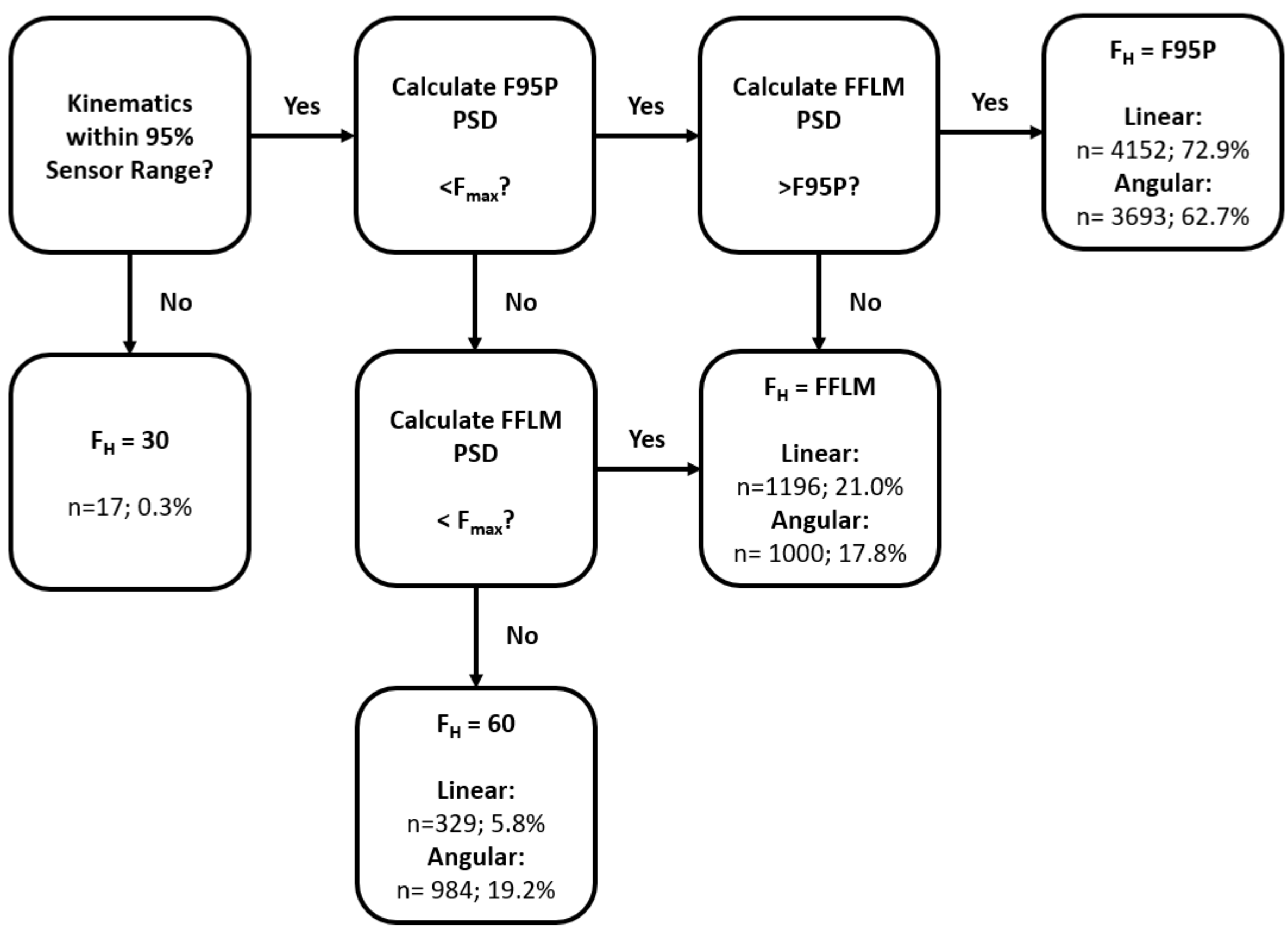

Figure 2. Flowchart illustrating the artefact attenuation method for selecting Frequency Class ($F_H$). F95P = frequency associated with 95$^{th}$ percentile PSD; FFLM = frequency associated with first local minima; $F_{max}$ = maximum 95$^{th}$ percentile PSD frequency for the head impact reconstruction laboratory impacts.

Median and outlier values for PLA and PAA were considerably higher when the CFC180 filter was applied to the raw kinematic signals in comparison to the Prevent Biometrics processed output ($p<0.01$), artefact attenuation method ($p<0.01$), Butterworth-200Hz ($p<0.01$) and head impact reconstruction laboratory kinematics (Fig. 3 and Supplementary B). PLA and PAA values as high as 294 g and 31.2 krad/$s^2$ were reported using the CFC180 filter in comparison to 101 g and 8.2 krad/$s^2$ from the Prevent Biometrics processed output. Both the artefact attenuation and Butterworth-200Hz filter produced similar PLA and PAA values within the magnitude range tested in the head impact reconstruction laboratory. The artefact attenuation and Butterworth-200Hz filter approach produced higher median PLA values than the Prevent Biometrics processed output (both $p<0.01$). The Butterworth-200Hz filter approach produced higher median PAA values than the Prevent Biometrics processed output ($p<0.01$) and outlier PAA values than the artefact attenuation method. The linear and angular acceleration pulse did not follow a half-sine/haversine shape when the CFC180 and Butterworth-200Hz filter was applied to certain HAE as high amplitude and frequency content were still present in the signal (Fig. 4).

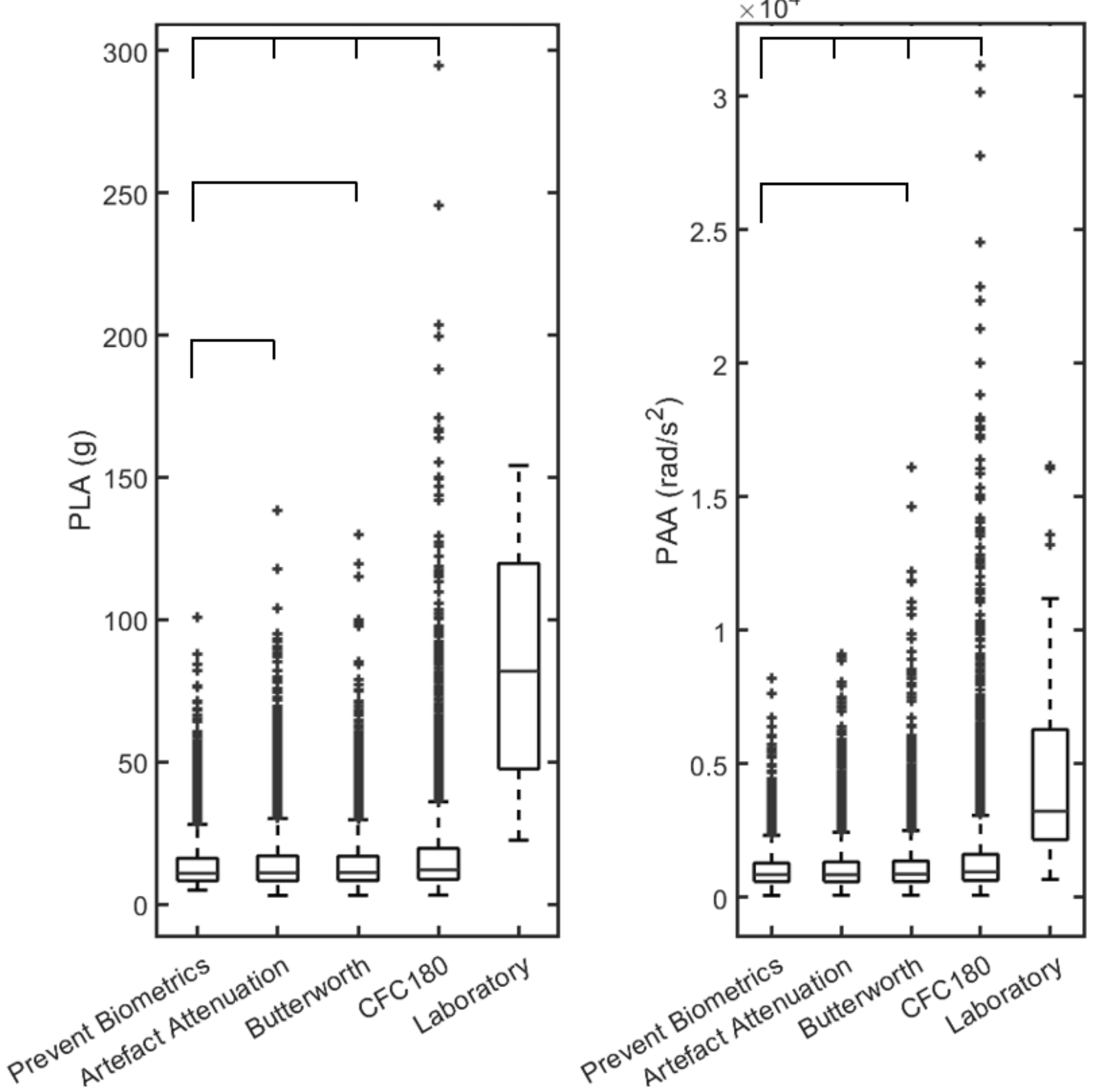

**Figure 3.** Peak linear acceleration (PLA) and peak angular acceleration (PAA) based on the Prevent Biometrics processed output, Artefact attenuation method, Butterworth-200Hz and CFC180 filter approach. Box plots indicate the median (box midline), interquartile range (box) and most extreme data points (whiskers) not considered outliers (crosses). Outliers are values greater than 1.5 times the IQR from top of the box. Horizontal lines above boxplots indicate statistical significance ($p<0.05$). Laboratory impacts were not included in the mixed linear effects models. Boxplots with outliers removed can be seen in Supplementary B.

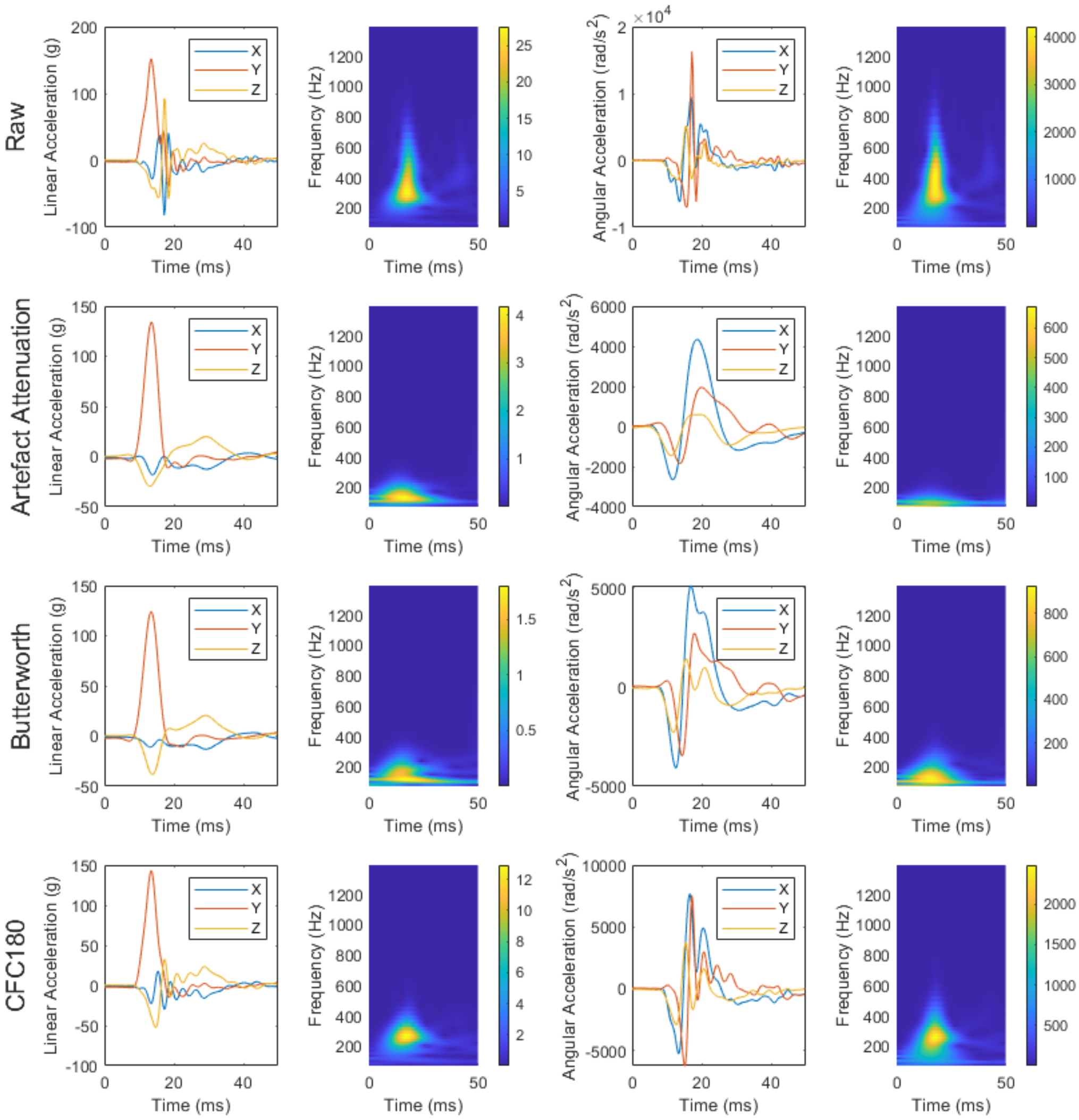

**Figure 4.** Linear and angular acceleration time-series with corresponding CWT scalogram on highest magnitude component for an example on-field HAE after artefact attenuation method, Butterworth-200Hz and CFC180 filter approach applied.

## DISCUSSION

An artefact attenuation filtering method for real-world/on-field iMG data has been developed based on the PSD characteristics of bare-headed laboratory impacts and on-field HAE. The rationale for the development of the artefact attenuation method was that peak kinematics differed considerably between different iMG systems used in similar rugby cohorts/environments (i.e. academy-level rugby league match play).[6] The artefact attenuation method produced an overall higher SNR than the Butterworth-200Hz and CFC180 filter in the head impact reconstruction laboratory environment and on-field PLA and PAA values within the magnitude range tested in the head impact reconstruction laboratory. The artefact attenuation method was designed as a fast, computationally inexpensive algorithm to be applied to any commercially available iMG system. Inter-study comparisons can be achieved through the use of the method's fundamental rigid body mechanics and signal processing principles (i.e. PSD) rather than machine learning approaches that require unique on-field training datasets and could be considered 'black box'.[29] Inter-study comparisons can allow head acceleration exposure and HAE mechanisms to be compared between sports. Additionally, a common signal processing approach means that datasets can be merged for analysis. This would be particularly useful for analyses that typically lack high sample sizes (e.g., kinematics of concussion injuries).[30]

Currently, the highest -3 dB cut-off frequency used by a commercially available iMG system is 300 Hz (CFC180 filter).[6] The artefact attenuation method enabled -3 dB cut-off frequencies as high as 500 Hz with the current dataset which is in line with the recommendations of Wu et al.[8] The cut-off frequency is also significantly below the excitation frequency of the skull, and thus complies with rigid body assumptions.[31] The median $F_H$ value of 60 Hz (indicative of -3dB cut-off frequency of 100 Hz) is low relative to other iMG system cut-off frequencies. However, this is likely due to the large number of indirect HAE (i.e., inertial head loading) in rugby HAE datasets which are lower frequency than direct HAE.[16] Conducting PSD on the impact pulse enables commonality for different iMG systems that utilise different output time-windows (usually range from 50 to 200 ms).[6,10] The SAE J211 impact test standard states that sensor sample rates should be a minimum of ten times $F_H$.[23] Therefore with an $F_{max}$ of 320 Hz, iMG sensors would need to sample at 3200 Hz as a minimum for the artefact attenuation method to measure short pulse duration HAE. Some iMG systems sample at 1000 Hz which has previously been highlighted as a concern.[8]

The use of a CFC180 filter without any artefact/noise attenuation processing resulted in the highest head kinematics for on-field data. This may explain why HAE of roughly 500 g and >50 krad/s$^2$ were reported by an iMG system utilising a 300 Hz cut-off frequency in Jones et al.[6] That is, high amplitude and frequency content remained in the signal after filtering and combined with numerical calculations (e.g., differentiation) and biomechanical transformations (e.g., transforming linear acceleration signal to the head CG) resulted in higher head kinematics being reported.[11] The significant differences in median and outlier PLA and PAA kinematics produced by the different filtering methods (Fig. 3) illustrate the need for a common approach to kinematic signal processing.

*Limitations*

There are biofidelity issues with a NOCSAE headform in comparison to a human head for impact reconstructions.[18] However, the maximum cut-off frequency based on the 95$^{th}$ percentile PSD frequency (312 Hz) equates to 520 Hz which is similar to that reported for cadaver-based head impact reconstructions.[8] There was also a characteristic haversine/half-sine pulse for the laboratory head impacts which are common in impact biomechanics,[32] and aligns with Equation 1. PSD analysis was conducted on an artificial raw angular acceleration pulse as described in the methodology. Ideally, PSD would be conducted on true raw angular accelerometer signals. Angular accelerometers are used by certain iMG systems.[6] $F_{max}$ was calculated based on bareheaded laboratory impacts. Although a padded impactor was utilised as part of the testing conditions, $F_{max}$ could be calculated for helmeted head impact reconstruction laboratory events and the artefact attenuation method assessed for helmeted contact sports (e.g., American Football). On-field ground truth data does not exist which makes validating the artefact attenuation method a challenge. However, the aim of the method is to allow inter-study comparison based on fundamental principles of head impact biomechanics and signal processing. Future work should assess the artefact attenuation method on datasets similar to Jones et al.[6] where multiple iMG systems are used in the same sport and playing-level to see if comparable kinematics are produced between iMG system. Validation could be assessed in controlled environments such as head impact reconstructions with cadavers equipped with iMG.[12] To enable field-based inter-study comparison, commonality/consensus is needed in terms of minimum sampling rates, sensor trigger and recording thresholds, head CG vector and calculation of parameters such as peak change in head angular velocity due to the high potential for non-zero head angular velocity values at the time of impact.[16] Additionally, commonality/consensus for what stages to apply the filter is important for comparable results. For laboratory data where true head kinematic signal is in the filter pass-band and noise in the stop-band, the filter only needs to be applied to the raw data before differentiation.[23] However in real-world datasets, artefact noise may be present in the transition-band (e.g., Fig. 1 and 4) and therefore only partially attenuated by filtering. Therefore, filtering after differentiation may be necessary as the amplitude of differentiated noise increases linearly with frequency.[11,17] Accordingly, the filter may also need to be applied after additional numerical calculations such as matrix transformations (e.g., converting from iMG to SAE J211 coordinate system for angular kinematics) and this requires further investigation.

Peak kinematics differ considerably between different iMG systems used in similar cohorts/environments. An artefact attenuation filtering method (artefact attenuation method) for real-world/on-field iMG data has been developed based on the PSD characteristics of bare-headed laboratory impacts. The artefact attenuation method is based on fundamental rigid body mechanics and signal processing principles. The method can be applied to all commercially available iMG kinematic signals with adequate sample rates to enable field-based inter-study comparisons. Ideally, the artefact attenuation method could contribute towards the development of a standard or consensus for processing sports-based head sensor kinematic signals that progresses in line with the state-of-the-art, similar to the automotive industry.

**CONFLICTS OF INTEREST**

GT has received research funding from Prevent Biometrics.

**ACKNOWLEDGEMENTS**

The authors would like to thank all players and club/team staff for participating in the study, World Rugby for supporting and funding the on-field data collection and Prevent Biometrics for their cooperation during the study such as providing raw data and sensor orientations. The authors would also like to thank Dr David Allen (Ulster University) for conducting the mixed linear effects model analysis.

**Supplementary A**. Time lapse of a perfectly elastic single rigid body head impacting a rigid surface modelled as a spring-mass system with derivation. Adapted from Tierney et al. [13]

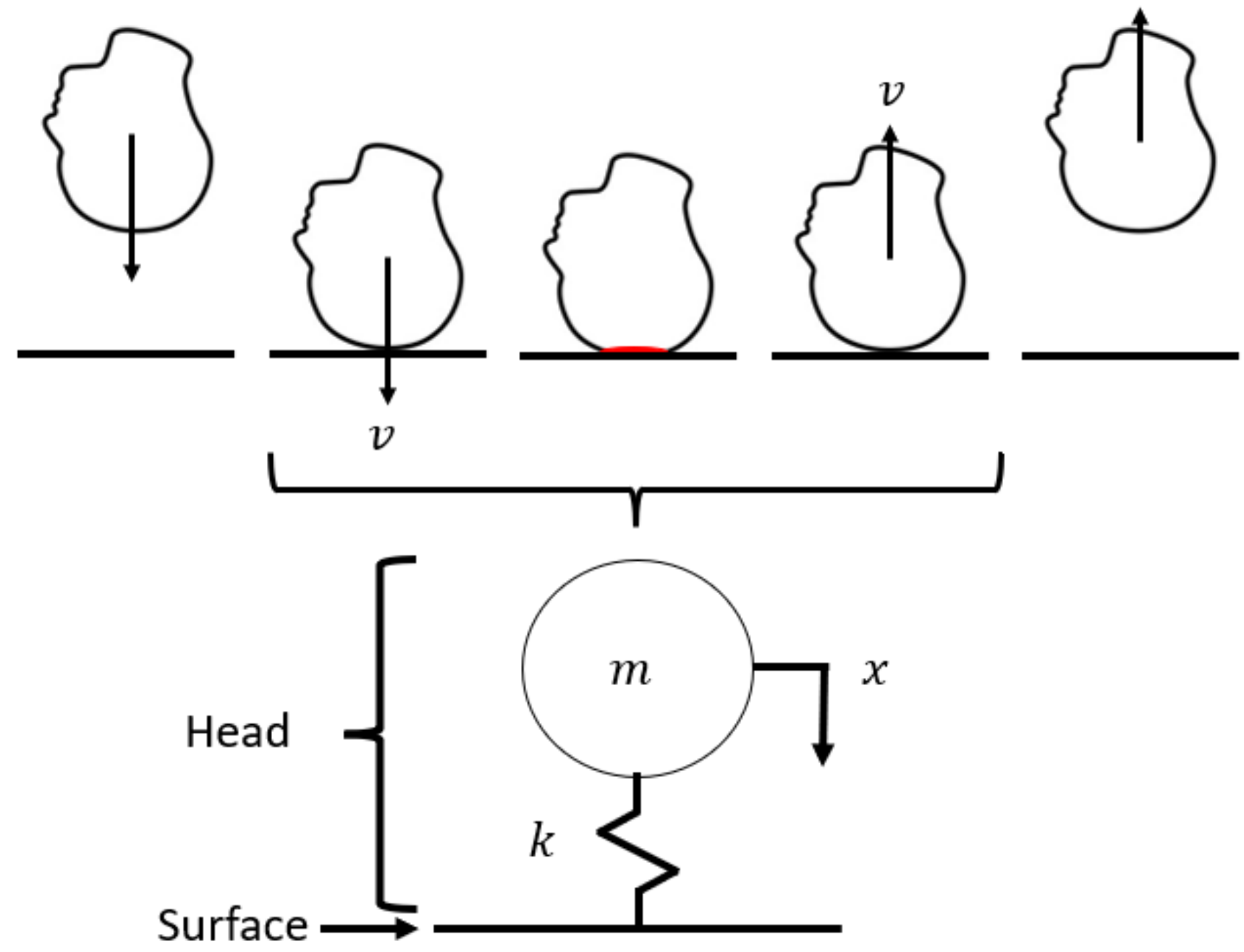

Spring mass system to represent head impact:

$$m\ddot{x}(t) + kx(t) = 0; \qquad \omega_n = \sqrt{\frac{k}{m}}$$

$$x(t) = \quad x(0)\cos(\omega_n t) \; + \; \frac{\dot{x}(0)}{\omega_n} \sin(\omega_n t)$$

for known $x(0) = 0$ and $\dot{x}(0) = v$;

$$x(t) = \frac{v}{\omega_n} \sin \omega_n t$$

$$\dot{x}(t) = v(t) = v_i \cos \omega_n t$$

$$\ddot{x}(t) = a(t) = -v_i \, \omega_n \sin \omega_n t = -v_i \sqrt{\frac{k}{m}} \sin \omega_n t$$

**Key:**
$m$ = Head Mass
$k$ = Head Stiffness
$\omega_n$= Head Natural Frequency
$v_i$ = Head Initial Velocity
$x$ = Head Displacement
$v$ = Head Velocity
$a$ = Head Acceleration

**Supplementary B.** Peak linear acceleration (PLA) and peak angular acceleration (PAA) based on the Prevent Biometrics processed output, Artefact attenuation method, Butterworth-200Hz and CFC180 filter approach with outliers removed.

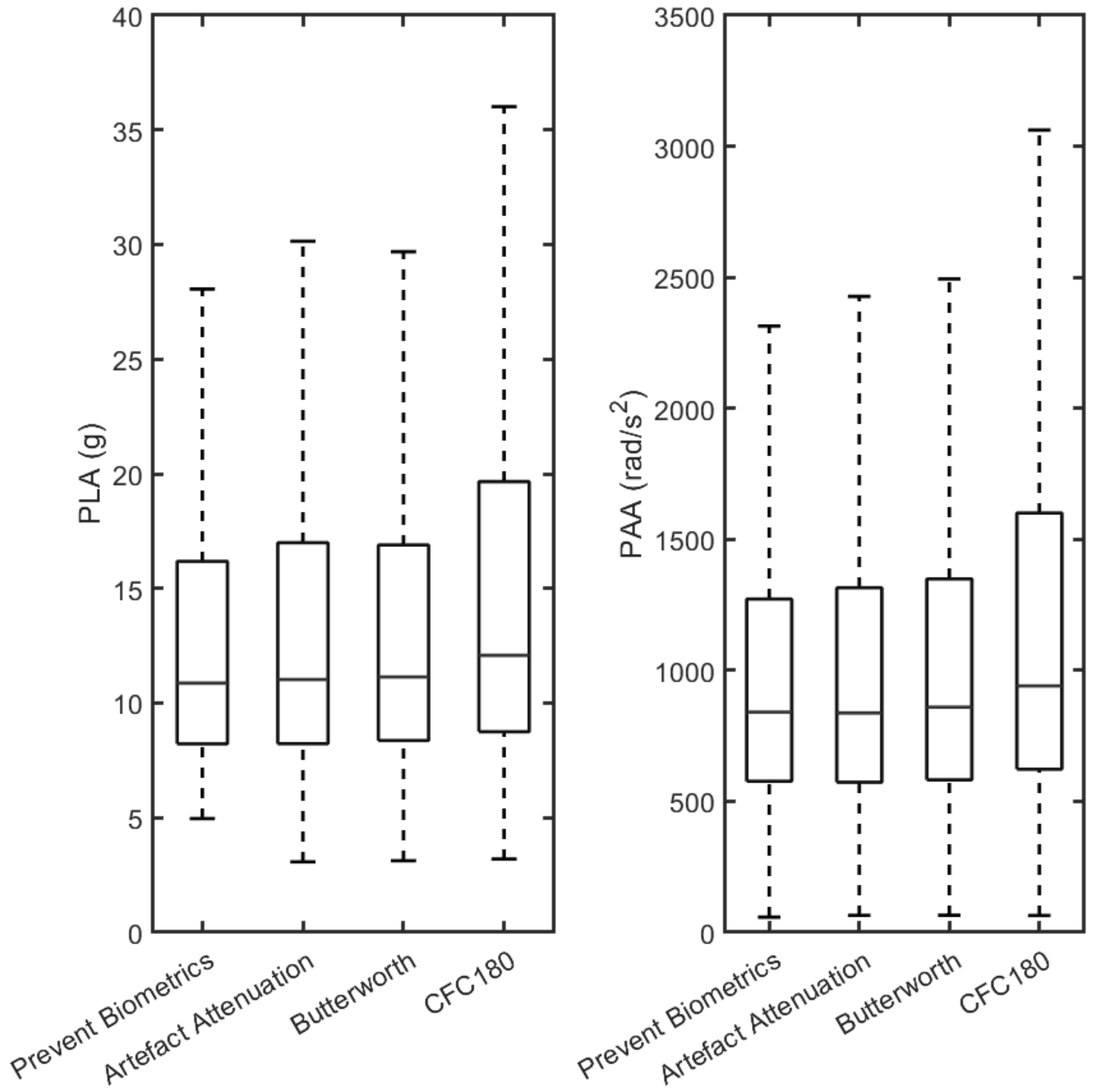